\documentclass[epsf,usegraphicx]{mn2e}

\title[Kepler Observations of V447 Lyr]{Kepler Observations of V447 Lyr: 
An Eclipsing U Gem Cataclysmic Variable}

\author[]
{Gavin Ramsay$^{1}$,  John K. Cannizzo$^{2,3}$, Steve B. Howell$^{4}$, 
Matt A. Wood$^{5}$, Martin Still$^{4,6}$, \and 
Thomas Barclay$^{4,6}$, Alan Smale$^{7}$\\
$^{1}$Armagh Observatory, College Hill, Armagh, BT61 9DG\\
$^{2}$CRESST and Astroparticle Physics Laboratory
NASA/GSFC, Greenbelt, MD 20771, USA\\
$^{3}$Department of Physics, University of Maryland,
Baltimore County, 1000 Hilltop Circle, Baltimore, MD 21250, USA\\
$^{4}$NASA Ames Research Center, Moffett Field, CA 94095, USA\\
$^{5}$Department of Physics and Space Sciences,
Florida Institute of Technology, Melbourne, FL\ \ 32901, USA\\
$^{6}$Bay Area Environmental Research Institute, Inc., 560
Third St. West, Sonoma, CA 95476, USA\\
$^{7}$NASA/Goddard Space Flight Center, Greenbelt, MD 20771, USA
}

\date{Accepted 2012 July 4. Received 2012 July 3; in original form 2012 June 20}

\begin{document}
\newcommand{\Msun} {$M_{\odot}$}
\newcommand{\Kepler}{\it Kepler}
\newcommand{\Porb}{P_{\rm orb}}
\newcommand{\nuorb}{\nu_{\rm orb}}
\newcommand{\eplus}{\epsilon_+}
\newcommand{\eminus}{\epsilon_-}
\newcommand{\cd}{{\rm\ c\ d^{-1}}}
\newcommand{\MdotL}{\dot M_{\rm L1}}
\newcommand{\Ldisk}{L_{\rm disk}}

\maketitle

\begin{abstract}
  We present the results of an analysis of {\Kepler} data covering 1.5
  years of the dwarf nova V447 Lyr. We detect eclipses of the
  accretion disk by the mass donating secondary star every 3.74 hrs
  which is the binary orbital period. V447 Lyr is therefore the first
  dwarf nova in the {\Kepler} field to show eclipses. We also detect
  five long outbursts and six short outbursts showing V447 Lyr is a U
  Gem type dwarf nova. We show that the orbital phase of the
  mid-eclipse occurs earlier during outbursts compared to quiescence
  and that the width of the eclipse is greater during outburst. This
  suggests that the bright spot is more prominent during quiescence
  and that the disk is larger during outburst than quiescence. This is
  consistent with an expansion of the outer disk radius due to the
  presence of high viscosity material associated with the outburst,
  followed by a contraction in quiescence due to the accretion of low
  angular momentum material. We note that the long outbursts appear to
  be triggered by a short outburst, which is also observed in the
  super-outbursts of SU UMa dwarf novae as observed using {\Kepler}.
\end{abstract}


\section{Introduction}

Cataclysmic Variable (CV) binary systems contain a white dwarf primary
star that accretes mass from a Roche lobe-filling late-type main
sequence secondary star.  Mass loss from the secondary through the
inner Lagrange point (L1) forms an accretion disk about the primary,
and viscosity within the accretion disk acts to transfer angular
momentum outward in radius allowing mass to migrate inward to the
surface of the white dwarf.  The disk and bright spot associated with
the accretion stream impact point are typically the brightest
components in a CV system (Warner 1995, Hellier 2001, Frank, King \&
Raine 2002).  The mean disk luminosity is ultimately provided by the
release of gravitational potential energy as the material migrates
inward through the disk given by $L_{\rm disk} \sim GM_1\dot M_1/R_1$,
where $\dot M_1$ is the mass transfer rate onto the primary of mass
$M_1$ and radius $R_1$.  The mass flow rate through L1 is governed by
the long-term evolution of the binary separation and the secondary
star itself, but the mass flow rate through the disk and onto the
primary is a function of the viscosity of the disk -- the higher the
viscosity, the higher the inward mass flow rate.

V447 Lyrae (KIC 8415928, $r$=18.4, Brown et al. 2011) is a
little-studied CV in the NASA {\Kepler} field of view. Its sky
co-ordinates are $\alpha=19^\textrm{h}00^{\rm m}19\fs92$ $\delta =
+44\degr 27'44\farcs9$ (2000.0).  It was discovered and announced as
GR 247 by Romano (1972) who noted a maximum photographic magnitude of
17.2 and a minimum fainter than 18.5. These observations were included
in Downes, Webbink \& Shara (1997) but the system was noted as
undetected in the 2MASS survey (Hoard et al. 2002) nor is it a known
X-ray source.

\begin{table*}
\begin{center}
\begin{tabular}{lllll}
\hline
Quarter & \multicolumn{2}{c}{Start} & \multicolumn{2}{c}{End} \\
        & MJD   & UT  & MJD & UT\\ 
\hline
Q6 (LC) & 55371.947 & 2010 Jun 24 22:46 & 55461.794 & 2010 Sep 22 19:04 \\
Q7 (LC) & 55462.673 & 2010 Sep 23 16:10 & 55552.049 & 2010 Dec 22 01:09 \\
Q8 (SC) & 55567.855 & 2011 Jan 06 20:42 & 55634.856 & 2011 Mar 14 20:15 \\
Q9 (SC) & 55641.007 & 2011 Mar 21 00:23 & 55738.434  & 2011 Jun 26 10:13 \\
Q10 (LC) & 55739.343 & 2011 Jun 27 08:16 & 55832.766 & 2011 Sep 28 18:24 \\
Q11 (SC) & 55833.696 & 2011 Sep 29 16:58 & 55930.837 & 2012 Jan 04 20:48 \\
\hline
\end{tabular}
\end{center}
\caption{Journal of Observations. The start and end MJD and UT dates are 
the mid-point  of the first and final cadence of the LC time series for each
  quarter respectively.}
\label{log}
\end{table*}

This work is the fifth in a series of publications focussing on the
CVs in the {\Kepler} mission field of view.  In Still et al. (2010), we
presented preliminary results for the periods observed in the Q2 data
for V344 Lyr, a previously little-studied CV in the {\Kepler} field.  In
Cannizzo et al. (2010), we presented results of the thermal viscous
disk instability model for CVs applied to the Q2--Q4 V344 Lyr outburst
time series data.  In Wood et al. (2011) we presented a detailed
analysis of the orbital and superhump periods present in the V344 Lyr
Q2--Q4 time series data.  In Cannizzo et al. (2012) we discussed the
outburst properties of V1504 Cyg and V344 Lyr over the first two years
of {\Kepler} observations and most recently (Barclay et al. 2012) we
report the serendipitous discovery of an SU UMa dwarf nova within 7
arcsec of a G-type star. Here we report {\Kepler} observations of V447
Lyr covering 1.5 years.

\section{Optical Spectroscopy}

\begin{figure}
\begin{center}
\setlength{\unitlength}{1cm}
\begin{picture}(12,6.3)
\put(-0.5,-0.5){\includegraphics{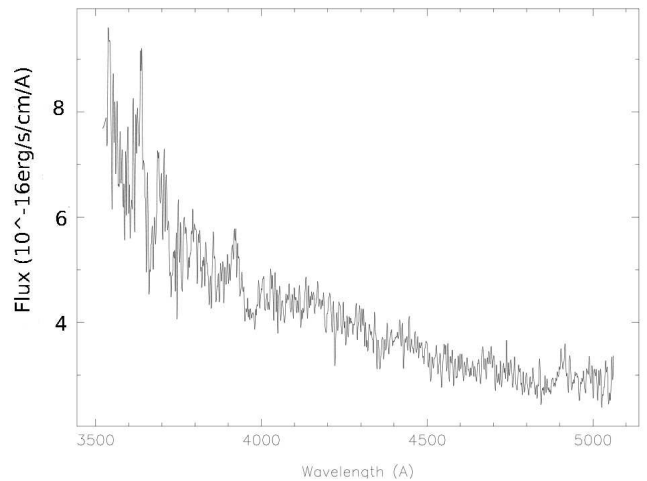}}
\end{picture}
\end{center}
\caption{An optical spectrum of V447 Lyr taken using the Palomar 200
  inch Telescope on 6 June 2011 when it was in a long outburst.}
\label{spectrum} 
\end{figure}

Figure \ref{spectrum} shows the blue spectrum of V447 Lyr obtained
using the double-beam spectrograph on the Mt. Palomar 200$^{"}$
telescope and confirms its CV nature. The spectrum was reduced in the
usual manner using IRAF 2-D and 1-D spectral reduction tools.  The
relative fluxes were provided by an observation of the
spectrophotometric standard star Feige 92 obtained 35 minutes prior to
the V447 Lyr spectrum. The night was not photometric, having thin to
moderate thick clouds passing by, thus the fluxes are approximate. The
red spectrum, obtained simultaneously, has low signal-to-noise but
shows that H$\alpha$ is also in emission.

This single 900 sec spectrum was recorded approximately two days into
a long outburst of V447 Lyr (LO4 in Table 2) and reveals complex line
profiles in the hydrogen lines. The lines of H$\beta$, H$\gamma$,
H$\delta$ and beyond show a broad absorption component with a slightly
off-center narrow emission line during this near-outburst-peak
spectrum.  In addition, a broad redshifted emission bump is observed,
the overall shape producing P Cyg-like line profiles. These line
profiles are similar to those observed in H$\alpha$ early into a
superoutburst of LL And (Howell \& Hurst 1996), with the complex
H$\alpha$ line structure quickly disappearing one night later.  The
optically thick wind outflow observed at this time has an average
velocity offset from the absorption line center of 3700 km/sec and a
velocity width (Full Width Zero Intensity) of 1200 km/sec, similar to
the line profiles and wind signatures observed in the ultraviolet
spectra of SW UMa, BC UMa, and TV Cor (Howell et al., 1995) during
superoutburst.

\section{Kepler Photometric Observations}

\begin{figure*}
\begin{center}
\setlength{\unitlength}{1cm}
\begin{picture}(12,6)
\put(-3.5,-3){\includegraphics{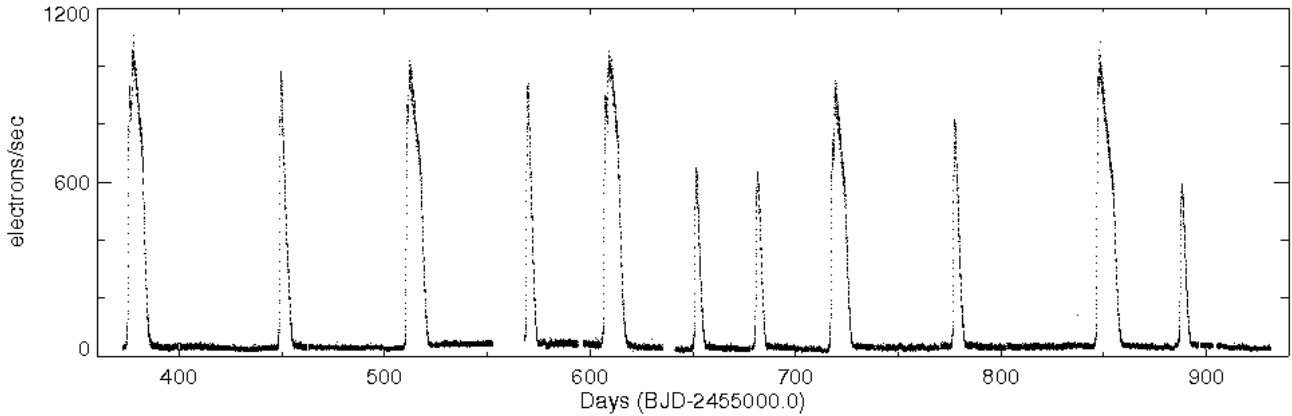}}
\end{picture}
\end{center}
\caption{Long Cadence {\Kepler} Q6--Q11 light curve of V447 Lyr in flux
  units. The time unit is in BJD - 2455000.0.} 
\label{long-cadence-light} 
\end{figure*}

The primary science objective of the {\Kepler} mission is to discover
Earth-sized planets in the habitable zone of Sun-like stars (Borucki
et al. 2010, Haas et al. 2010). The spacecraft is in an Earth-trailing
orbit allowing it a continuous view of the target field over the
planned 3.5-yr mission lifetime (recently extended for at least a
further two years).  The shutterless photometer has a 116 deg$^2$
field of view and makes use of 6.54-s integrations, but only pixels
containing pre-selected targets are saved due to bandwidth and memory
constraints.  Bandwidth limits impose that only up to 170,000 targets
can be observed in {\it long cadence} (LC) mode, where 270
integrations are summed for an effective 28.4-min exposure, and up to
512 targets can be observed in {\it short cadence} (SC) mode, where 9
integrations are summed for an effective 58.8-s exposure.  Gaps in the
{\Kepler} data streams result from, for example, 90$^\circ$ spacecraft
rolls every 3 months (called quarters), monthly data downloads using
the high-gain antenna as well as unplanned safe-mode and loss of fine
point events.  For further technical details see Haas et al. (2010),
Koch et al. (2010) and Caldwell et al. (2010).

{\Kepler} data are provided as quarterly FITS files by the Science
Operations Center after being processed through the standard data
reduction pipeline (Jenkins et al. 2010).  After the raw data are
corrected for bias, shutterless readout smear, and sky background,
time series are extracted using simple aperture photometry (SAP).  The
start and end times of each quarter of {\Kepler} data which are used in
this study are shown in Table \ref{log}. (We note that when SC mode
data are obtained, LC are also obtained).

\begin{table}
\begin{center}
\begin{tabular}{llrrr} 
\hline
Outburst & Quarter/ & Start Date & Duration & Amplitude \\
         & Cadence  &            & (Days)   & (mag) \\
\hline
 LO1 & Q6/LC  &374.3 & 11.8 & 3.5 \\
 SO1 & Q6/LC  &448.1 & 7.3 & 3.8\\
 LO2 & Q7/LC  &509.6 & 12.4 & 3.8 \\
 SO2 & Q8/SC  &568.3 & 6.5 & 3.2\\
 LO3 & Q8/SC  &605.5 & 12.7 & 3.5\\
 SO3 & Q9/SC  &650.1 & 6.4 & 3.5\\
 SO4 & Q9/SC  &679.7 & 6.5 & 3.2\\
 LO4 & Q9/SC  &716.1 & 11.8 & 4.0\\
 SO5 & Q10/LC  &776.1 & 6.6 & 3.3\\
LO5  & Q11/SC& 845.7 & 13.0 & 3.5 \\
SO6  & Q11/SC& 886.3 & 6.1 & 2.9\\
\hline
\end{tabular}
\end{center}
\caption{A summary of outbursts observed in {\Kepler} data of V447 Lyr.
  LO indicates a `long' outburst and SO a `short' outburst. The start
  date is the time of the rise to outburst where the date is BJD -
  2455000.0. The duration of the outburst is given in days and the
  amplitude in mag.}
\label{out-log}
\end{table}

We show the long cadence light curve for V447 Lyr obtained over
quarters Q6--11 in Figure \ref{long-cadence-light}.  The light curve
shows 11 outbursts: 6 of these have a `short' duration clustered
around a mean of 6.6 days and 5 have a `long' duration clustered
around a mean of 12.3 days (Table \ref{out-log}).  The ratio of the
duration of the long outbursts to short outbursts is 1.9 and is
consistent with that seen in systems with long orbital periods (and
hence high mass transfer rates, e.g. Warner 1995). Indeed, it is very
similar to that seen in U Gem ($P_{orb}$=4.25 hrs) whose short and
long outbursts have a typical duration of 5 and 12 days respectively
(Cannizzo, Gehrels \& Mattei 2002).  The mean recurrence time between
successive long outbursts is 118 days (the mean recurrence time in U
Gem is $\sim$120 days, Szkody \& Mattei 1984). There are only two
successive short outbursts and they were separated by 30 days.

However, perhaps the most notable aspect of the light curve is that an
eclipse is observed every 3.74 hrs (Figure \ref{LO5-curve}). This
feature is seen because the binary inclination angle is high enough to
cause the physically larger mass-donating secondary star to eclipse
the white dwarf (if the inclination is close to 90$^{\circ}$), or the
accretion disc and bright spot (if the inclination is slightly less
than 90$^{\circ}$), once every orbital period leading to an apparent
dimming in the light from the system. We show below that the likely
inclination of this system is such that the white dwarf itself is not
eclipsed. We therefore identify V447 Lyr as the first eclipsing dwarf
nova in the {\Kepler} field.

\section{A precursor to the long outbursts}

\begin{figure}
\begin{center}
\setlength{\unitlength}{1cm}
\begin{picture}(12,5.4)
\put(0,0){\includegraphics{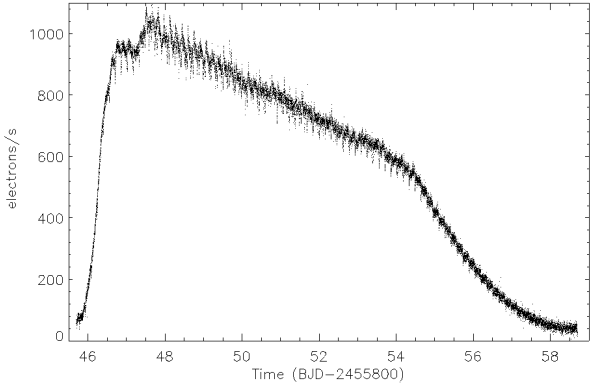}}
\end{picture}
\end{center}
\caption{The light curve of long outburst 5 (see Table
  \ref{out-log}), showing an eclipse of the accreting white dwarf by
  the late-type main sequence star every 3.74 hrs.}
\label{LO5-curve} 
\end{figure}

The short duration outbursts show a sharp rise followed by an
exponential type decay. In contrast, the long outbursts all show an
initial sharp rise (as in the short outbursts) followed by a plateau
lasting less than a day, and then an increase in flux (see Figure
\ref{LO5-curve}), followed by a gradual decline in flux. This
`precusor' to the super-outburst has been seen in all of the
super-outbursts of SU UMa CVs observed using {\Kepler} (e.g. Still et
al. (2010), Barclay et al.  (2012)). A normal (or short) outburst is
thought to trigger a super (or long) outburst when the material in the
outer radii of the accretion disk becomes ionised.

This precursor has not been seen in long outbursts of U Gem type CVs
until very recently (Cannizzo 2012). We attribute this to the fact
that the cadence of observations has not been high enough to resolve
this precursor. However, given a precursor has been seen in super
outbursts and now in long outbursts, we suggest this precursor is
common to long outbursts from accreting CVs in general. Any model
aiming to reproduce long/super outbursts will have to account for this
feature.

\section{Eclipsing Timings}

We initally estimated the time of the mid-eclipse in individual
eclipses during each of the outbursts which were observed in SC
mode. Given that the exposure time of each SC data point was nearly
one minute and the eclipse profile was rather noisy, the uncertainty
on the eclipse times were relatively large. However, it allowed us to
obtain a working eclipse ephemeris which we used to phase the SC light
curve.  To improve the signal-to-noise of the eclipse profile we split
up the light curve into sections consisting of 3 orbital cycles when
the system was in outburst and up to 30 cycles during quiescence and
phase folded and binned the data. This allowed us to refine the linear
eclipse ephemeris so that during an outburst the eclipse mid-point was
defined as $\phi$=0.0. This gave the eclipse ephemeris:\newline

\begin{equation}
T_{o} = BJD 2455569.4134(2) + 0.1556270(1) E \noindent
\end{equation}

where the numbers in parentheses give the standard error on the last
digits. 

\section{Folded light curves}

\begin{figure}
\begin{center}
\setlength{\unitlength}{1cm}
\begin{picture}(8,5.2)
\put(-0.5,-0.2){\includegraphics{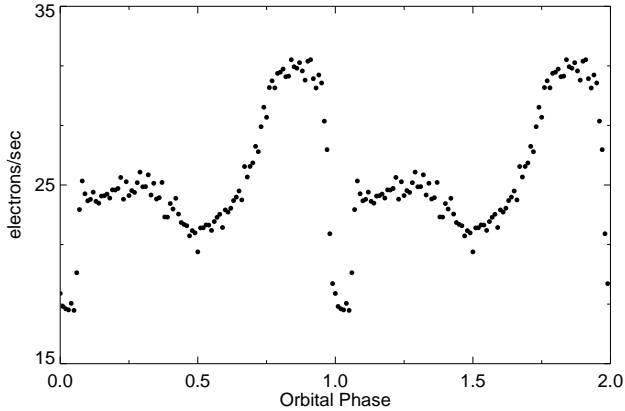}}
\end{picture}
\end{center}
\caption{The folded light curve obtained from data taken during the
  quiescent interval MJD=55687.0--55714.0, ie between outbursts SO4
  and LO4 (see Table 2).}
\label{low-fold} 
\end{figure}

\begin{figure*}
\begin{center}
\setlength{\unitlength}{1cm}
\begin{picture}(12,10)
\put(-4,-0.5){\includegraphics{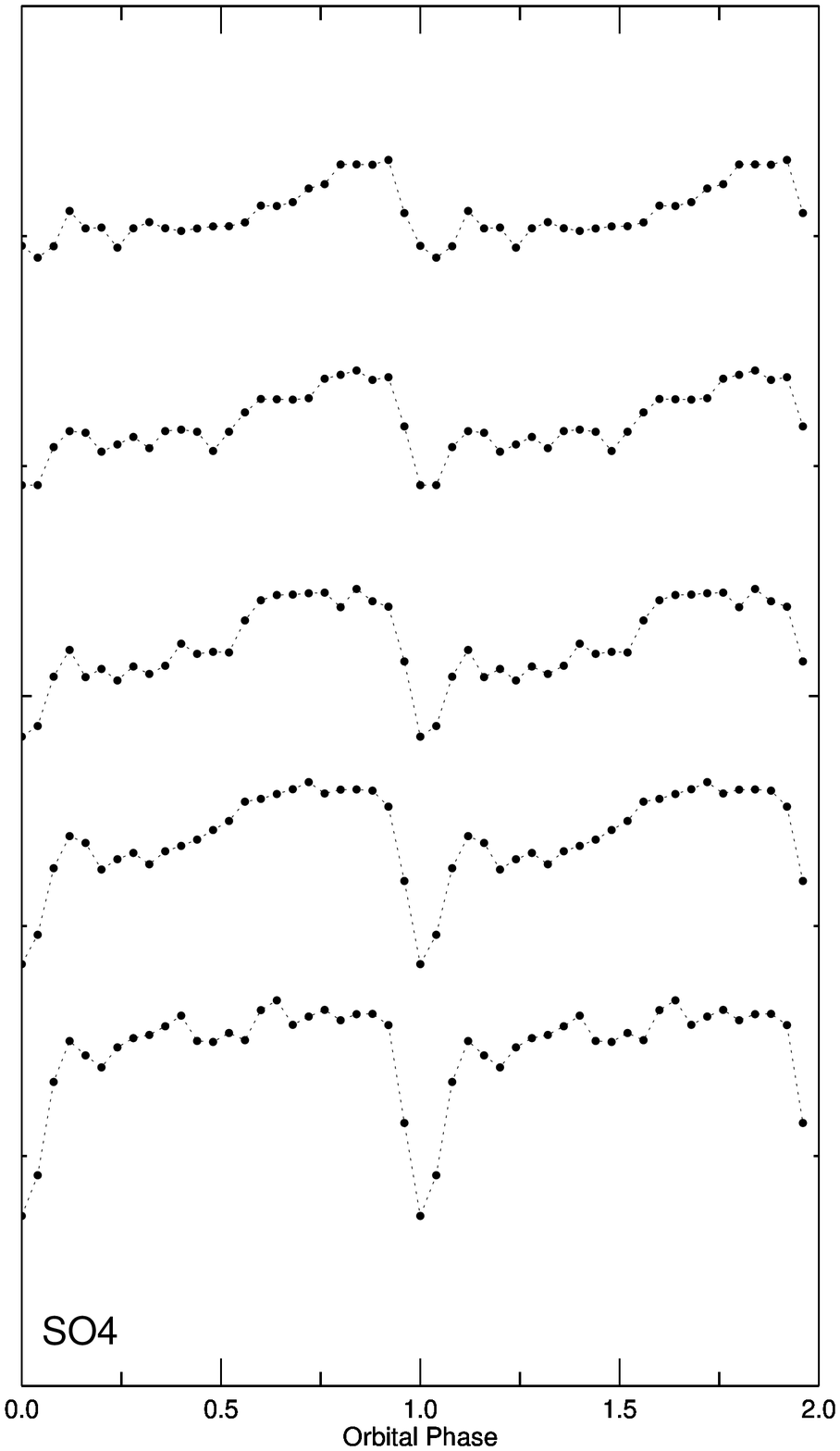}}
\put(2,-0.5){\includegraphics{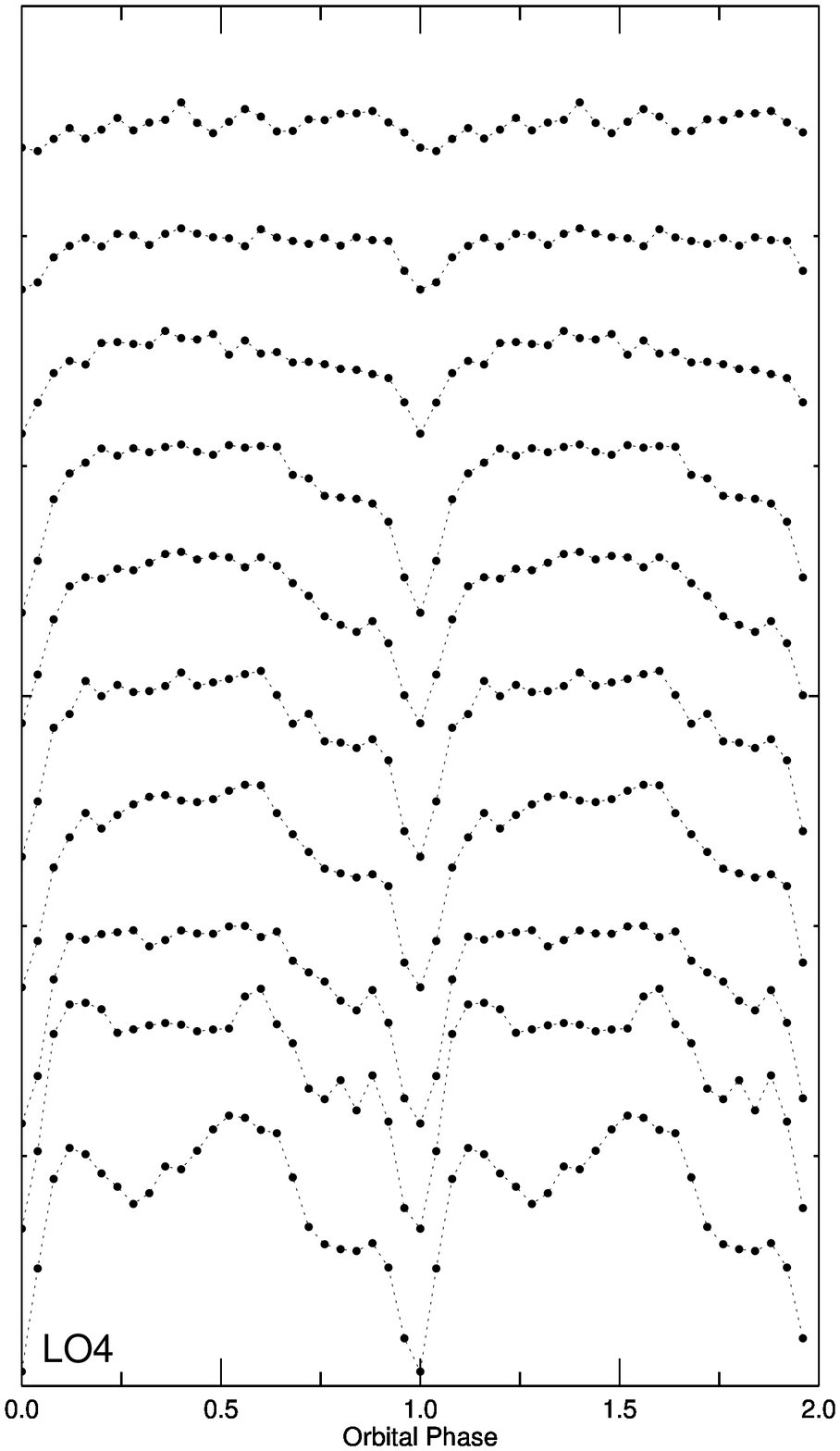}}
\put(8,-0.5){\includegraphics{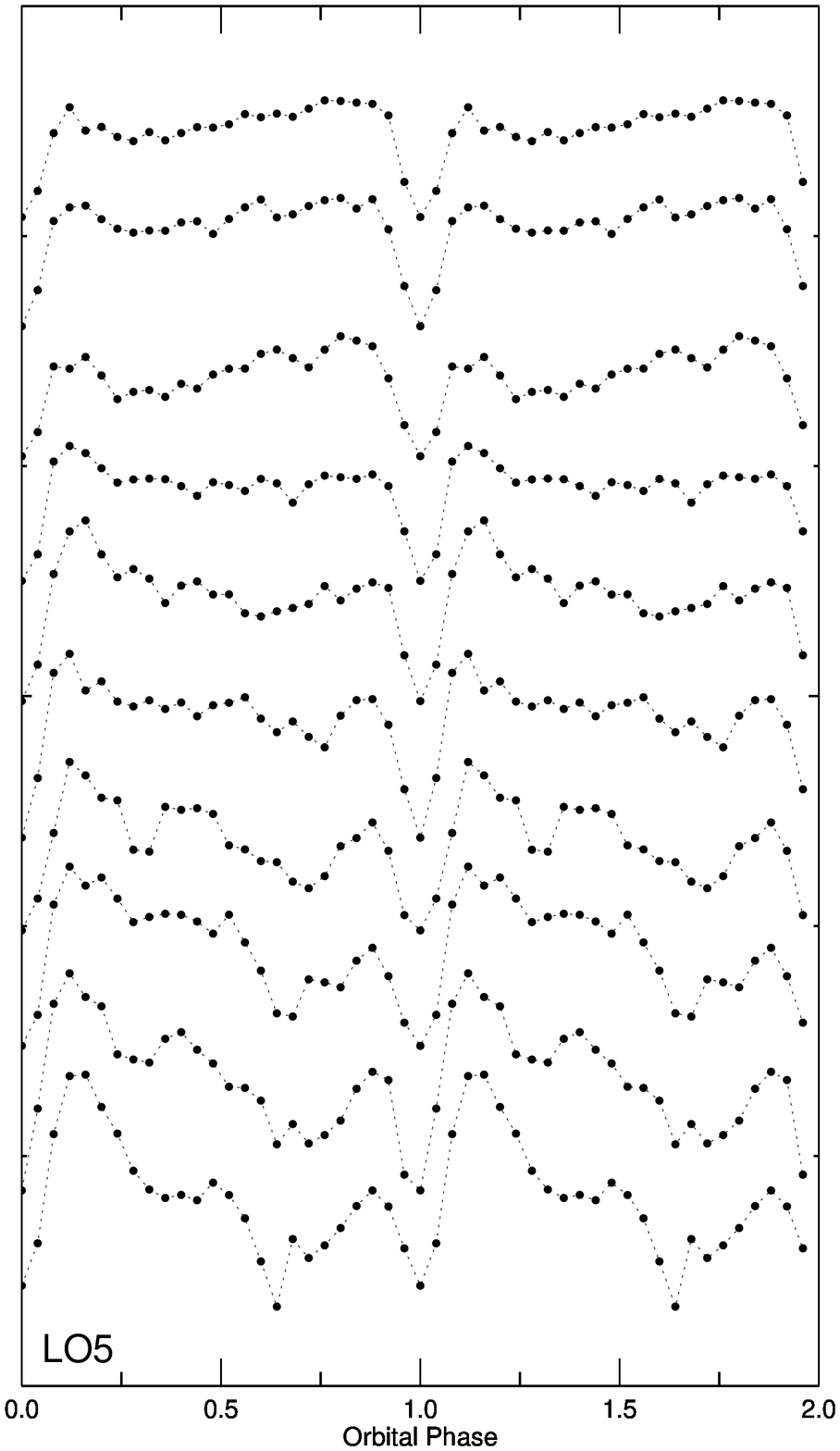}}
\end{picture}
\end{center}
\caption{We show a set of folded light curves from a short outburst
  (SO4 in the left hand panel) and two long outbursts (L04, middle
  panel and LO5, right hand panel). We have de-trended the light curve
  from each outburst and split the light curves into 0.7 day sections
  and then folded and binned each on the eclipse ephemeris -- time
  increases upwards.}
\label{folded} 
\end{figure*}

When we fold the light curve derived from intervals of quiescence we
find a peak in the folded light curve at $\phi\sim$0.8--0.9 (Figure
\ref{low-fold}) which is due to the bright spot where the accretion
stream hits the accretion disk and is seen approximately face-on at
this phase angle (see Wood et al. 1986 for a breakdown of the
predicted contribution from the white dwarf, accretion disk and bright
spot as a function of orbital phase). The full width at half maximum
of the eclipse profile is $\Delta\phi\sim$0.08. Assuming we can derive
the mass of the secondary star ($M_{2}=0.27$) from the mass-period
relationship (Patterson et al. 2005) and a mass of $M_{1}=0.6$ for the
white dwarf ($q=$0.45) we predict an inclination angle $i=80^{\circ}$
(or $i=82.5^{\circ}$ for $M_{1}=0.8$), using the Figure 2 of Horne
(1985).

In the left hand panel of Figure \ref{folded} we show a set of folded
light curves taken from a short outburst (SO4), where we have
detrended the light curve to remove the effect of the decline from
outburst, and have split the light curve into sections 0.7 days in
length and then folded the light curve according to the ephemeris
derived in the previous section. At outburst maximum the brightness
out of eclipse is more uniform, and as the system approaches
quiescence, a peak appears in the folded light curve at $\phi$=0.8 due
to the bright spot becoming progressively more prominent as the system
approaches quiescence. During outburst the bright spot makes a
relatively significantly smaller contribution to the optical light
from the system compared to quiescence.

In the middle panel of Figure \ref{folded} we show the results from a
similar analysis of a long outburst (LO4). We find that at outburst
maximum (the lowermost curve in Figure \ref{folded}), there is a peak
in the folded light curve at $\phi\sim$0.55. In contrast, in the right
hand panel of Figure \ref{folded} we show the results of the same
analysis for the next long outburst (LO5). At maximum brightness there
is a prominent peak in the folded light curve shortly after the
eclipse, while there is a minumum in the light curve at
$\phi\sim$0.6. Over the course of the outburst the brightness out of
eclipse becomes more uniform (the analysis for L03 is very similar to
LO5). We speculate that this evolution of the accretion disc (and the
differences between the LO4 and that of LO3 and LO5) is due to the
evolution of sprial shocks in the accretion disc as observed in U Gem
(Groot 2001).

\section{The eclipse profile}

\begin{figure*}
\begin{center}
\setlength{\unitlength}{1cm}
\begin{picture}(12,11)
\put(-3,-0.4){\includegraphics{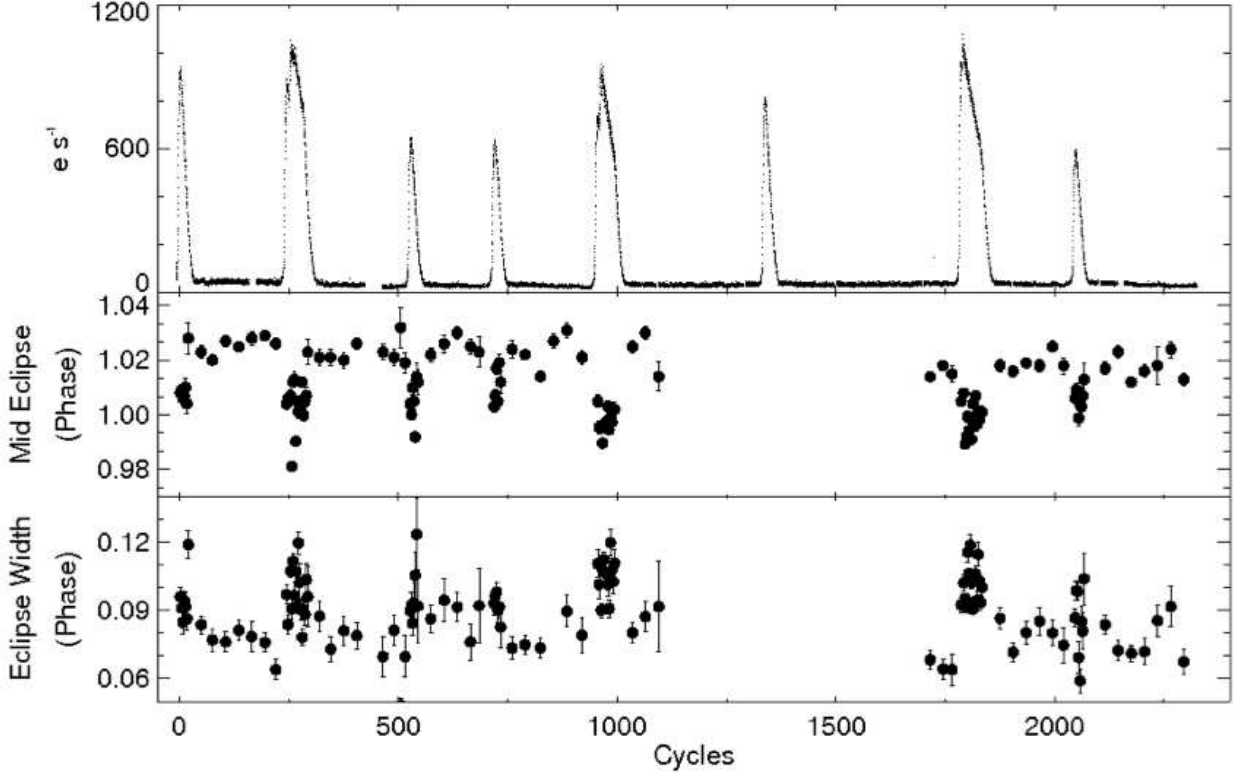}}
\end{picture}
\end{center}
\caption{In the top panel we show the light curve derived from the LC
  mode data. In the middle panel we plot the phase of the mid-eclipse
  where $\phi$=0.0 has been defined as the mid-eclipse time during
  outburst. The lower panel shows the eclipse width in units of
  orbital phase. (The results for the middle and lower panels were
  derived using SC data. The gap results from only LC data being
  obtained during the corresponding quarter).}
\label{phase-width} 
\end{figure*}

We now turn our attention to the characteristics of the eclipse
profile. As indicated in \S 5, we extracted three successive cycles of
data during outbursts and phase folded and binned these data using the
eclipse ephemeris. For intervals of quiescence we did this for up to
30 orbital cycles of data. We fitted the eclipse profiles using a
model consisting of a linear trend plus a Gaussian profile and
determined the mid-point and width of the eclipse. We show the results
of this in Figure \ref{phase-width}.

The most obvious result is that during an outburst the phase of the
mid-eclipse is centered near $\phi$=0.0 (since that defined our
ephemeris), whereas the phase of the mid-eclipse during quiescence
occurs at later phases ($\phi$=1.02--1.03). There is some suggestion
that after a long outburst the phase of eclipse during short outbursts
gets progressively later. During long outbursts the full width half
maximum of the eclipse is greater compared to quiescence and also
compared to short outbursts. The difference between the results from
quiescence and outbursts is largely due to the relative contribution
of the bright spot which is seen strongly in the folded light curve
from quiescence (Figure \ref{low-fold}) but not during outburst
(Figure \ref{folded}). If it was only the accretion disk which was
being eclipsed then it would only be the eclipsed width which would
change and not the eclipse phase.

\begin{figure*}
\begin{center}
\setlength{\unitlength}{1cm}
\begin{picture}(6,9)
\put(-3.5,0.5){\includegraphics{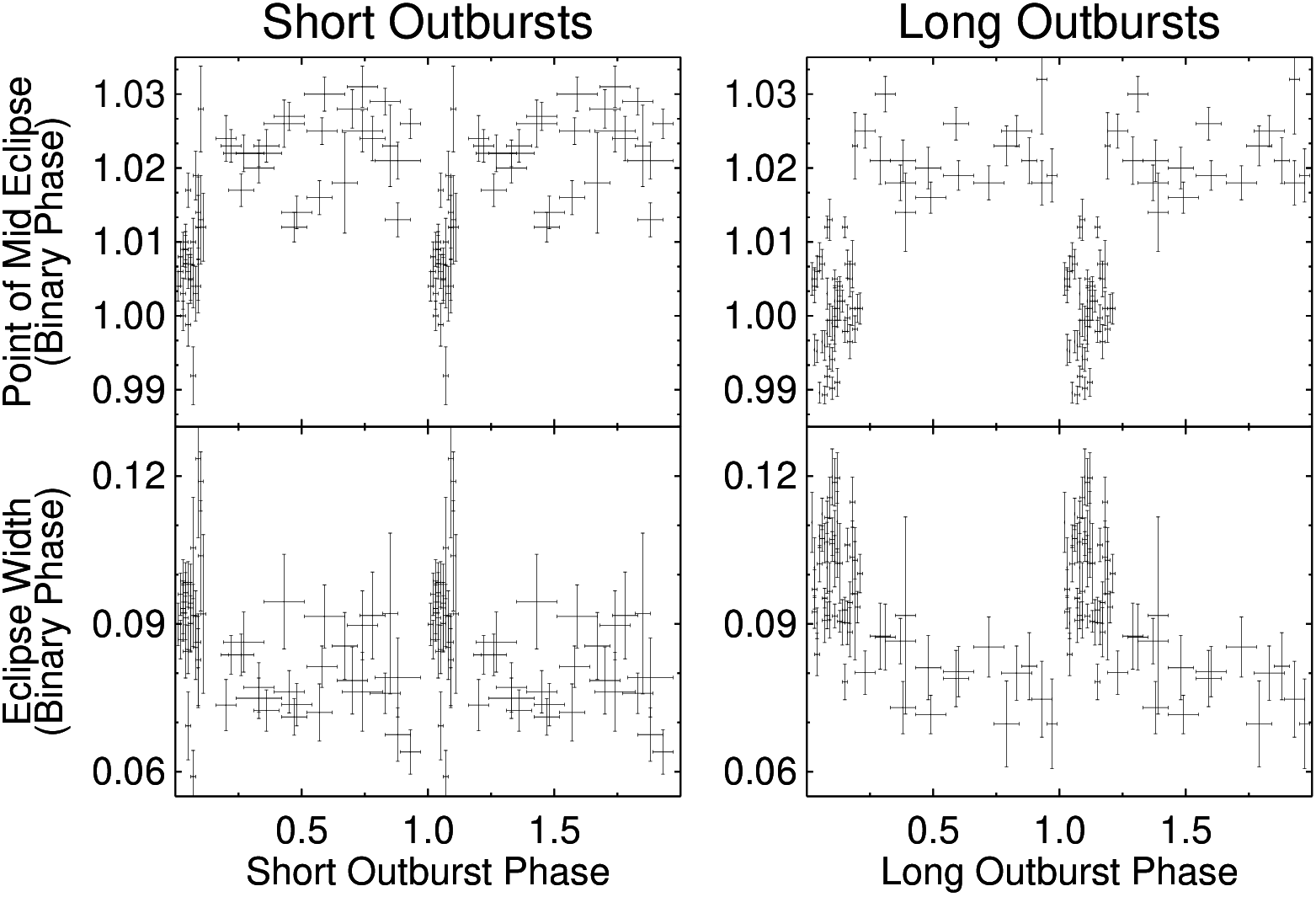}}
\end{picture}
\end{center}
\caption{We phased outbursts such that $\phi$=0.0 corresponded to the
  start of the outburst and $\phi$=1.0 corresponded to the point just
  before the start of the next outburst. We then folded and binned the
  results for the mid-eclipse times and eclipse widths shown in Figure
  \ref{phase-width} to show how these parameters vary over the course
  of a short outburst (left panels) and a long outburst (right
  panels).\vspace{3mm}}
\label{NO-SO} 
\end{figure*}

To obtain a better overview of the differences between the short and
long outbursts, we took each outburst and phased it such that
$\phi$=0.0 defined the start of the outburst, with the end phase being
the start of the next burst. We then folded the points of mid-eclipse
and the eclipse width for short and long bursts - we show the results
of this in Figure \ref{NO-SO}. This shows that there is a tendancy for
the orbital phase of mid-eclipse to be earlier (and with a larger
eclipse width) in long outbursts compared to short outbursts, which we
interpret as the accretion disk has a tendancy to be larger during a
long outburst compared to a short outburst. In short outbursts there
is some evidence that the point of mid-eclipse increases after the
outburst followed by a decrease, which may suggest a decrease in the
size of the disk followed by an increase leading up to the next
outburst. During a long outburst the width of the eclipse tends to
decrease after outburst. We explore the physical causes for these
variations in \S \ref{discuss}.

\section{Searching for Additonal periods}

We searched for periodic signals (other than the orbital period) in
the short cadence data. We did this by splitting up the data into
sections which were taken from quiescent intervals and also individual
outbursts, where we detrended the data to remove the effect of the
general decline from outburst. The Lomb Scargle periodogram was used
to obtain power spectra for each data section.

Power spectra taken from the quiescent data showed a strong signal at
a period very close (within the error) to the orbital
period. Similarly, power spectra of detrended light curves during
outbursts also showed maximum power at a period consistent with the
orbital period. We therefore found no significant evidence for
super-humps in these data.  The detection of super-humps in U Gem
(Smak \& Waagen 2004) is rather controversial (Schreiber 2007).
However, we note that the novalike variable BB Dor has an orbital
period only slightly shorter than V447 Lyr at 3.70 hr
(Rodr{\'{\i}}guez-Gil et al. 2012), and is reported to display
superhumps with a period of 3.93 hr. Continued monitoring of V447 Lyr
to search for evidence of super-hump behaviour is therefore
recommended since systems with an orbital period just shorter than 4
hrs may define the edge of the superhump instability strip in period
space.

\section{Discussion}
\label{discuss}

The cadence of our SC observations preclude us from resolving distinct
binary components (e.g. white dwarf and accretion disc) in the eclipse
profile. However, we have found a systematic difference in the orbital
phase of the mid-eclipse as seen in quiescence and in outburst. We
therefore caution that unless you can detect the eclipse of the white
dwarf, mid-eclipse times determined from quiescence should not be
mixed with times determined from outburst.

Detailed observations from eclipsing dwarf novae place strong
constraints on the accretion disk radius as a function of time. Two
well-studied systems are U Gem (Smak 1984) and Z Cha (O'Donoghue
1986). The disk expands dramatically during outburst, and then
contracts exponentially.  Anderson (1988) presents a simple disk model
in which the sudden appearance of high viscosity material in outburst
causes the rapid expansion, and the accretion of low angular momentum
material from the secondary causes the slower contraction. Figure 1 of
Anderson shows a collection of all the U Gem data, which shows a
variation of $r_d/a$ between about 0.29 and 0.37. Thus the outburst
radius expands by $\sim$30\% during outburst. Anderson infers a
minimum mass in the outer accretion disk of
$\sim10^{-9}$\Msun. Ichikawa \& Osaki (1992) examined the radius
expansion question using a detailed numerical model for the accretion
disk thermal limit cycle mechanism and confirmed Anderson's overall
results.

Although the variation in the eclipse width shown in the bottom panel
of Figure \ref{phase-width} shows more variability than the results of
Smak (1984), it is similar quantitatively to U Gem - which is
apparently a near twin of V447 Lyr in many ways.  It is more difficult
with our data to state clearly what the form of the decay is, due to
the scatter in the quiescent data, but in going from quiescence to
outburst the disk expands from $\sim$0.08 to $\sim$0.12 in phase
units, an increase of $\sim$50\%. Given the similar orbital periods of
V447 Lyr and U Gem, a fit using a simple model like Anderson's should
also produce a lower limit disk mass $\sim10^{-9}$\Msun.

\section{Conclusions}

We report observations of V447 Lyr which show that it is the first
dwarf nova in the {\Kepler} field to show eclipses. It has an orbital
period of 3.74 hrs and shows almost equal numbers of long and short
outbursts, which makes it a near twin of the well studied dwarf nova U
Gem.  By fitting the mean eclipse profile of three successive eclipses
we find that the phase of the mid-eclipse occurs earlier during
outbursts compared to quiescence and that the width of the eclipse is
greater during an outburst. This suggests that the accretion disk has
a larger radius during outburst compared to during quiescence and is
consistent with an expansion of the outer disk radius due to the
presence of high viscosity material associated with the outburst,
followed by a contraction in quiescence due to the accretion of low
angular momentum material. {\Kepler} observations of dwarf novae
outbursts have found that super-outbursts in the shorter orbital
period dwarf novae appear to be triggered by a normal outburst. We
find that long outbursts also appear to be triggered by short
outbursts in V447 Lyr. This indicates that this is a general phenomena
found in CVs which any outburst model will have to reproduce.

\section{Acknowledgments}

{\Kepler} was selected as the 10th mission of the Discovery Program.
Funding for this mission is provided by NASA, Science Mission
Directorate.  All of the data presented in this paper were obtained
from the Multimission Archive at the Space Telescope Science Institute
(MAST).  STScI is operated by the Association of Universities for
Research in Astronomy, Inc., under NASA contract NAS5-26555. Support
for MAST for non-HST data is provided by the NASA Office of Space
Science via grant NAG5-7584 and by other grants and contracts. This
material is based upon work supported by the National Science
Foundation under Grant No. AST 1109332 to the Florida Institute of
Technology. Armagh Observatory is supported by the Northern Ireland
Executive through the Dept of Culture Arts and Leisure.


\vspace{-8mm}
\begin{thebibliography}{}

\bibitem{}Anderson, N. 1988, ApJ, 325, 266
\bibitem{}Barclay, T., Still, M.,
  Jenkins, J. M., Howell, S. B., Roettenbacher, R. M., 2012, MNRAS,
  422, 1219
\bibitem{}Borucki, W.~J., et  al.\ 2010, Science, 327, 977
\bibitem{}Brown, T. M., Latham, D. W.,
  Everett, M. E., Esquerdo, G. A., 2011, AJ, 141, 112
\bibitem{}Caldwell, D., et al.\ 2010, ApJ, 713, L92
\bibitem{}Cannizzo, J.~K., 2012, ApJL, submitted
\bibitem{}Cannizzo, J.~K., Gehrels,  N., Mattei, J. A., 2002, ApJ, 579, 760
\bibitem{} Cannizzo, J.~K., Still, M. D., 
Howell, S. B., Wood, M. A., Smale, A. P., 2010, ApJ, 725, 1393  
\bibitem{} Cannizzo, J.~K., Smale, A. P., 
Wood, M. A., Still, M. D., Howell, S. B., 2012, ApJ, 747, 117 
\bibitem{} Downes, R., Webbink, 
	R.~F., \& Shara, M.~M.\ 1997, PASP, 109, 345 
\bibitem{} Frank, J., King, A., 
	\& Raine, D.~J.\ 2002, Accretion Power in Astrophysics, by Juhan Frank
	and Andrew King and Derek Raine, pp.~398.~ISBN 0521620538.~Cambridge,
	UK: Cambridge University Press, 2002.
\bibitem{}Groot, P., 2001, ApJ, 551, L89
\bibitem{}Haas, M. R., et al.\ 2010, ApJ, 713, L115
\bibitem{}Hellier, C. 2001,
    Cataclysmic Variable Stars: How and Why They Vary,
	Springer-Praxis Books in Astronomy \& Space Sciences: Praxis
	Publishing
\bibitem{} Hoard, D.~W., Wachter, 
S., Clark, L.~L., \& Bowers, T.~P.\ 2002, ApJ, 565, 511 
\bibitem{}Horne, K., 1985, MNRAS, 213, 129
\bibitem{}Howell, S. B., Szkody, P., 
Sonneborn, G., Fried, R., Mattei, J., Oliversen, R. J., Ingram, D.,
Hurst, G. M., 1995, ApJ, 453, 454
\bibitem{}Howell, S. B., Hurst,    G. M., JBAA, 106, 29
\bibitem{}Ichikawa, S., \& Osaki, Y. 1992, PASJ, 44, 15
\bibitem{} Jenkins, J.~M., et al.\  2010, ApJL, 713, L87 
\bibitem{} Koch, D.~G., et al.\ 2010, ApJ, 713, L79 
\bibitem{}O'Donoghue, D. 1986, MNRAS, 220, 23P
\bibitem{}Patterson, J., et al.\ 2005, PASP, 117, 1204 
\bibitem{}Rodr{\'{\i}}guez-Gil, P., Schmidtobreick, L., Long, K.~S., 
et al.\ 2012, MNRAS, 422, 2332
\bibitem{} Romano, G.\ 1972, Information 
	Bulletin on Variable Stars, 645, 1 
\bibitem{}Schreiber, M. R.,  2007, A\&A, 466, 1025
\bibitem{}Smak, J. 1984, Acta Astr., 34, 93
\bibitem{}Smak, J., \& Waagen, E. O. 2004, Acta Astron., 54, 433
\bibitem{} Still, M., Howell, S.~B., 
	Wood, M.~A., Cannizzo, J.~K., \& Smale, A.~P.\ 2010, 
	ApJ, 717, L113
\bibitem{}Szkody, P., Mattei, J. A., 1984, PASP, 96, 988
\bibitem{} Warner, B. 1995, Cataclysmic Variable
	Stars (Cambridge: Cambridge)
\bibitem{}Wood, J., Horne, K., Berriman,
G., Wade, R., O'Donoghue, D., Warner, B., 1986, MNRAS, 219, 629
\bibitem{}Wood, M. A., Still, M., Howell, S.~B.,
Cannizzo, J.~K., \& Smale, A.~P., 2011, ApJ, 741, 105
\end{thebibliography}
\end{document}